\documentclass[11pt,a4paper]{article}
\usepackage[T1]{fontenc}
\usepackage[utf8]{inputenc}
\usepackage{lmodern}
\usepackage[margin=2.4cm]{geometry}
\usepackage{amsmath,amssymb,bm,mathtools}
\usepackage{graphicx,booktabs,xcolor}
\usepackage[numbers,sort&compress]{natbib}
\usepackage{authblk}
\usepackage[hidelinks]{hyperref}

\newcommand{\vr}{\mathbf{r}}
\newcommand{\vR}{\mathbf{R}}
\newcommand{\vs}{\mathbf{s}}
\newcommand{\vq}{\mathbf{q}}
\newcommand{\vk}{\mathbf{k}}
\newcommand{\vp}{\mathbf{p}}
\newcommand{\vE}{\mathbf{E}}
\newcommand{\vD}{\mathbf{D}}
\newcommand{\vn}{\hat{\mathbf{n}}}
\newcommand{\teps}{\bm{\epsilon}}            % relative permittivity tensor
\newcommand{\tepsi}{\bm{\epsilon}^{-1}}
\newcommand{\rot}{\mathsf{O}}                % rotation matrix
\newcommand{\Id}{\mathsf{I}}
\newcommand{\tr}{\operatorname{tr}}
\newcommand{\enorm}[1]{|#1|_{\epsilon}}      % metric norm
\newcommand{\dete}{\det\teps}
\newcommand{\Qe}{\mathbb{Q}^{\epsilon}}

\title{Electrostatics of homogeneous anisotropic dielectrics:\\
Green's function, metric-adapted multipoles, and consequences for buried charges,
surface probes and inclusions}

\author{Jo\~{a}o Elias F. S. Rodrigues\thanks{\texttt{rodrigues.joaoelias@esrf.fr}; ORCID 0000-0002-9220-5809}}
\affil{European Synchrotron Radiation Facility (ESRF), 71 Avenue des Martyrs, 38000 Grenoble, France}
\date{}

\begin{document}
\maketitle

\begin{abstract}
Most crystals, textured ceramics and layered van der Waals materials are dielectrically anisotropic, yet the textbook treatment of electrostatics in matter is almost always restricted to isotropic media. We give a compact, self-contained treatment of electrostatics in a linear, homogeneous, non-magnetic medium described by a symmetric, positive-definite permittivity tensor $\teps$, valid in the quasistatic regime. The Green's function is obtained in one step through the metric $\vr\cdot\tepsi\vr$ and is used to formulate the Dirichlet and Neumann problems in terms of the conormal derivative $\vn\cdot\teps\nabla$. Four consequences are worked out. (i) The Coulomb interaction is non-central and is screened along a principal axis by the geometric mean of the two \emph{other} principal permittivities, so that the interaction is least screened along the axis of largest permittivity. (ii) Beyond the dipole order, the multipole moments must be adapted to the metric: the conventional traceless quadrupole tensor gives errors of order 50--100\% in monoclinic CdWO$_4$, and the correct, $\teps$-adapted quadrupole tensor is given. (iii) For a planar grounded electrode or an insulating boundary, an exact point image exists for an arbitrary orientation of the crystal axes, located at an ``$\epsilon$-mirror'' point. One consequence is that a collinear four-probe measurement on the surface of an anisotropic conductor gives $(\det\bm{\rho})^{1/2}(\vn\cdot\bm{\rho}\,\vn)^{-1/2}$. (iv) A charge buried below the surface of an anisotropic dielectric is seen from outside at an apparent depth and lateral position that depend on the crystal face. In rutile TiO$_2$ the apparent depth ranges from $0.72h$ on (001) to $1.39h$ on (110), and on (011) the apparent position is shifted laterally by $0.26h$. Finally, the Laplace problem of an anisotropic ellipsoidal inclusion in an anisotropic host under a uniform field is solved. The resulting depolarization tensor corrects the field inside spherical pores in rutile by up to 16\% and the Maxwell Garnett permittivity of porous, textured rutile by up to 6\%.
\end{abstract}


%==========================================================================
\section{Introduction}\label{sec:intro}
Anisotropy, the dependence of a physical property on direction, is the rule rather than the exception in crystalline solids. Equilibrium properties such as the dielectric permittivity, the magnetic susceptibility, elasticity and piezoelectricity, as well as transport properties such as electrical and thermal conductivity, are described by tensors whose form is fixed by the point-group symmetry of the crystal \cite{nye1985physical,haussuhl2007physical,kocks2000texture}. For a symmetric rank-two tensor, cubic crystals have one independent component, tetragonal, hexagonal and trigonal crystals have two, and orthorhombic, monoclinic and triclinic crystals have three, four and six, respectively \cite{nye1985physical,haussuhl2007physical}.

Despite this, the standard treatment of electrostatics in matter (Green's functions, boundary-value problems, multipole expansions, images, inclusions in uniform fields) is developed almost exclusively for isotropic media \cite{Jackson1998,Greiner1998,griffiths2012introduction,landau1995electrodynamics}. Anisotropic electrostatics has been addressed in more specialised contexts: the effective response of anisotropic inclusions and composites \cite{Helsing1995,Levy1997,Giordano2006,Giordano2008,sihvola1999electromagnetic,Milton2002}, Hall and conductivity tensors of composites \cite{Christiansson1997,Novikov2006}, Green's functions and boundary-element methods for anisotropic solids and piezoelectrics \cite{Baghai-Wadji1997,Fischerauer1997,PanA2000,PanB2000,Tonon2001,Shiah2012,Thurieau2012}, image theory for anisotropic half-spaces and layered media \cite{Lindell1997,Tan2002,Hanninen2004,Mele2001}, charge storage in rutile \cite{Damamme2008}, wave propagation in anisotropic and hyperbolic media \cite{Chen1982,chen1983theory,Seoktae2010,Potemkin2012}, and the optics of crystals and anisotropic multilayers probed by generalised ellipsometry \cite{BornWolf1999,Schubert1996,Schubert1998}. Recently, the screened Coulomb interaction in thin films and two-dimensional (2D) semiconductors, the Rytova--Keldysh problem \cite{Keldysh1979}, has been generalised to anisotropic permittivity tensors \cite{Galiautdinov2019,Galiautdinov2026}. This is motivated by excitons in anisotropic 2D crystals such as phosphorene \cite{Rodin2014,Prada2015} and by the strongly anisotropic dielectric response of layered materials such as hexagonal boron nitride (hBN) and the transition-metal dichalcogenides \cite{Laturia2018}. Deng \cite{Deng2020} showed that the response of an anisotropic film to an external charge is conveniently described by volume and surface polarisation charges.

The aim of this work is twofold. First, we give a compact and pedagogically transparent derivation of the Green's function, the Green identities, the Dirichlet and Neumann representations, and the multipole expansion for a homogeneous anisotropic dielectric. All of these follow from a single ingredient: the metric $\vr\cdot\tepsi\vr$ induced by the permittivity tensor. In doing so we correct a point that is easily overlooked: beyond the dipole order the multipole moments must be \emph{adapted to the metric}, and the conventional traceless quadrupole tensor does not give the correct far field. Second, and this is the main motivation of the present version, we work out physical and practical consequences that can be compared with experiment:
the non-central and direction-dependent Coulomb interaction between point defects (Sec.~\ref{sec:green});
exact images for planar electrodes and insulating boundaries with arbitrarily oriented crystal axes, and the four-probe resistance of anisotropic conductors (Sec.~\ref{sec:bvp});
the face-dependent apparent depth and lateral displacement of a charge buried in a crystal, relevant for Kelvin-probe and electrostatic-force microscopy and for the scanning-electron-microscope mirror method (Sec.~\ref{sec:surface});
and the solution of the Laplace problem of an anisotropic inclusion in an anisotropic host under a uniform field, with its consequences for the effective permittivity of porous textured ceramics (Sec.~\ref{sec:inclusion}).
Cubic NaCl, tetragonal rutile TiO$_2$ and monoclinic CdWO$_4$ are used as examples throughout.

%==========================================================================
\section{Model and range of validity}\label{sec:model}

\subsection{Meaning of ``anisotropic medium''}
We consider a medium whose \emph{macroscopic} response is local, linear and homogeneous, $\vD=\epsilon_0\teps\vE$, where $\teps$ is a constant, dimensionless $3\times3$ tensor. Two physically different situations lead to this description.
(a) \emph{Intrinsic (crystalline) anisotropy}: a single crystal of non-cubic symmetry, for which $\teps$ is fixed by the point group \cite{nye1985physical}.
(b) \emph{Textural or form anisotropy}: an aggregate of isotropic or anisotropic constituents arranged with a preferred orientation, for example textured polycrystals, aligned fibres or platelets, laminates and porous ceramics with oriented pores \cite{kocks2000texture,sihvola1999electromagnetic,Milton2002,Rodrigues2020}. In case (b), $\teps$ is an \emph{effective} tensor, and the results below hold on length scales much larger than the characteristic size of the microstructure. Section~\ref{sec:inclusion} shows explicitly how such an effective tensor arises from the constituents.

In both cases we assume that $\teps$ is real, symmetric and positive definite. Symmetry follows from thermodynamic arguments in the absence of a static magnetic field and of magnetoelectric coupling \cite{nye1985physical,landau1995electrodynamics}. Gyrotropic (Hall-like) media with an antisymmetric part of the tensor, hyperbolic media with principal values of opposite sign \cite{Potemkin2012} and spatially dispersive media are excluded. Their Green's functions are qualitatively different. The notation is as follows: $\mathbf{a}\cdot\mathbf{b}$ is the scalar product, $\teps\mathbf{a}$ the matrix--vector product, $\mathsf{M}^{\top}$ the transpose of a matrix $\mathsf{M}$, and $\Id$ the $3\times3$ identity. We also define the metric norm
\begin{equation}
\enorm{\vR}\equiv\left(\vR\cdot\tepsi\vR\right)^{1/2}.\label{eq:norm}
\end{equation}

\subsection{Quasistatic regime and frequency-dependent tensors}
The analysis is strictly electrostatic. For time-dependent fields it applies in the quasistatic regime, i.e.\ when the relevant dimensions $\ell$ of the system are much smaller than the wavelength in the medium, $\ell\ll\lambda/n$, so that retardation can be neglected. In this regime $\teps$ may be replaced by the frequency-dependent tensor $\teps(\omega)$, which is complex for absorbing media. The formulas below then remain valid by analytic continuation, with the principal branch of the square roots, as long as the real part of $\teps$ is positive definite. Close to polar-phonon resonances some principal components may become negative, and the medium is then hyperbolic \cite{Potemkin2012}. The tensor $\teps(\omega)$ of a crystal or a thin film is measured by (generalised) spectroscopic ellipsometry and infrared reflectivity, which are analysed with the $4\times4$ matrix formalism for plane waves in arbitrarily anisotropic layered systems \cite{BornWolf1999,Schubert1996,Schubert1998,Jellison2011,Rodrigues2020}. The present results are the long-wavelength ($\vk\to0$) counterpart of that formalism and use the same tensor as input. For the static examples below we use the room-temperature values
\begin{equation}
\teps_{\rm NaCl}=5.6\,\Id,\quad
\teps_{\rm TiO_2}=\begin{pmatrix}89&0&0\\0&89&0\\0&0&173\end{pmatrix},\quad
\teps_{\rm CdWO_4}=\begin{pmatrix}8.5&1.2&0\\1.2&9.5&0\\0&0&7.7\end{pmatrix},\label{eq:tensors}
\end{equation}
for NaCl (space group $Fm\bar3m$) \cite{ashcroft1976solid,nye1985physical}, rutile ($P4_2/mnm$, $z\parallel c$) \cite{Diebold2003,raeyTiO2,Parker1961} and CdWO$_4$ ($P2/c$) \cite{Jellison2011}. The CdWO$_4$ tensor is written in a Cartesian frame that is not the principal frame. Its principal values are $7.7$ (twice) and $10.3$, and its in-plane principal axes are rotated by $33.7^\circ$ with respect to $x$ and $y$. It serves as an example of a tensor with off-diagonal components.

\subsection{Field equations}
With $\nabla\cdot\vD=\rho$, $\nabla\times\vE=\mathbf{0}$ and $\vE=-\nabla\Phi$, the potential obeys the anisotropic Poisson equation
\begin{equation}
\nabla\cdot\left(\teps\nabla\Phi\right)=-\rho/\epsilon_0 .\label{eq:poisson}
\end{equation}
The Green's function is defined by
\begin{equation}
\nabla\cdot\left(\teps\nabla G(\vr,\vr_0)\right)=-\delta(\vr-\vr_0)/\epsilon_0,\qquad
\Phi(\vr)=\int_\Omega G(\vr,\vr_0)\,\rho(\vr_0)\,d^3r_0 ,\label{eq:Gdef}
\end{equation}
where the integral representation holds in unbounded space with $\Phi\to0$ at infinity. Boundaries are treated in Sec.~\ref{sec:bvp}.

%==========================================================================
\section{Green's function and the anisotropic Coulomb interaction}\label{sec:green}

\subsection{Derivation}
Let $\teps^{1/2}$ be the unique symmetric positive-definite square root of $\teps$. If $\rot\in\mathsf{SO}(3)$ diagonalises the tensor, $\rot\teps\rot^{\top}=\mathrm{diag}(\epsilon_1,\epsilon_2,\epsilon_3)$, then $\teps^{1/2}=\rot^{\top}\mathrm{diag}(\epsilon_1^{1/2},\epsilon_2^{1/2},\epsilon_3^{1/2})\rot$. We introduce the stretched coordinates
\begin{equation}
\vs=\teps^{-1/2}\vr,\qquad \nabla_{r}=\teps^{-1/2}\nabla_{s}.\label{eq:stretch}
\end{equation}
This combines in one step the rotation to the principal axes and the rescaling of each axis by $\epsilon_i^{-1/2}$. Then $\nabla_r\cdot\teps\nabla_r=\nabla_s\cdot\teps^{-1/2}\teps\,\teps^{-1/2}\nabla_s=\Delta_s$, and $\delta(\vr-\vr_0)=\delta\big(\teps^{1/2}(\vs-\vs_0)\big)=(\dete)^{-1/2}\delta(\vs-\vs_0)$. Equation~\eqref{eq:Gdef} becomes an isotropic Poisson equation,
\begin{equation}
\Delta_s G=-(\dete)^{-1/2}\,\delta(\vs-\vs_0)/\epsilon_0 .\label{eq:Gs}
\end{equation}
With the Fourier representation $G=\int\tilde G(\vq)\,e^{i\vq\cdot(\vs-\vs_0)}d^3q/(2\pi)^3$, where $\vq$ is the wave vector conjugate to $\vs$, one finds $\tilde G(\vq)=(\dete)^{-1/2}/(\epsilon_0 q^2)$. The inverse transform is the Coulomb potential $1/(4\pi|\vs-\vs_0|)$ \cite{arfken2012mathematical,byron1992mathematics}. Since $|\vs-\vs_0|^2=(\vr-\vr_0)\cdot\tepsi(\vr-\vr_0)$,
\begin{equation}
\boxed{\,G(\vr,\vr_0)=\frac{1}{4\pi\epsilon_0\,(\dete)^{1/2}\,\enorm{\vr-\vr_0}}\,}\,.\label{eq:G}
\end{equation}
Equivalently, in terms of a wave vector $\vq'$ conjugate to $\vr$, $\tilde G(\vq')=1/(\epsilon_0\,\vq'\cdot\teps\vq')$. This form makes explicit that $\teps$ enters only through the quadratic form $\vq'\cdot\teps\vq'$. Equation~\eqref{eq:G} agrees with the results obtained by coordinate transformations in Refs.~\cite{Fischerauer1997,sihvola1999electromagnetic,Giordano2008}. We checked numerically for the CdWO$_4$ tensor that Eq.~\eqref{eq:G} satisfies Eq.~\eqref{eq:Gdef} away from $\vr_0$ and that the flux of $\vD$ through a closed surface around $\vr_0$ equals the enclosed charge (relative deviation $<10^{-5}$).

\subsection{Field, displacement and force}
For a point charge $Q$ at the origin, with $\vR=\vr$,
\begin{equation}
\Phi=\frac{Q}{4\pi\epsilon_0(\dete)^{1/2}\enorm{\vR}},\quad
\vE=\frac{Q\,\tepsi\vR}{4\pi\epsilon_0(\dete)^{1/2}\enorm{\vR}^3},\quad
\vD=\frac{Q\,\vR}{4\pi(\dete)^{1/2}\enorm{\vR}^3}.\label{eq:pointfields}
\end{equation}
The equipotentials are the ellipsoids $\vR\cdot\tepsi\vR=\mathrm{const}$, whose semi-axes are proportional to $\epsilon_i^{1/2}$ along the principal axes. The displacement field is radial, but the electric field is parallel to $\tepsi\vR$, i.e.\ normal to the ellipsoids. The force exerted by charge $Q_2$ on charge $Q_1$,
\begin{equation}
\mathbf{F}_{12}=\frac{Q_1Q_2\,\tepsi\vR}{4\pi\epsilon_0(\dete)^{1/2}\enorm{\vR}^3},\qquad \vR=\vr_1-\vr_2,\label{eq:force}
\end{equation}
is therefore not central. For a uniaxial crystal with $\kappa=\epsilon_\parallel/\epsilon_\perp$, the maximum angle $\theta_{\max}$ between $\mathbf{F}_{12}$ and $\vR$ satisfies $\tan\theta_{\max}=(\kappa-1)/(2\sqrt{\kappa})$. In rutile ($\kappa=1.94$) this gives $\theta_{\max}=18.7^\circ$. Because $\mathbf{F}_{12}$ and $\mathbf{F}_{21}$ are equal and opposite but not collinear, a bound pair such as a donor--acceptor pair or a vacancy--interstitial complex experiences an orientational torque $\bm\tau=\vR\times\mathbf{F}_{12}$. The interaction energy can be written as $U=Q_1Q_2/[4\pi\epsilon_0\,\epsilon_{\rm eff}(\hat{\vR})\,|\vR|]$ with the direction-dependent screening
\begin{equation}
\epsilon_{\rm eff}(\hat{\vR})=(\dete)^{1/2}\left(\hat{\vR}\cdot\tepsi\hat{\vR}\right)^{1/2},\qquad
\epsilon_{\rm eff}(\hat{\mathbf{e}}_i)=\sqrt{\epsilon_j\epsilon_k}\quad(i,j,k\ \text{distinct principal axes}).\label{eq:epseff}
\end{equation}
Along a principal axis, the Coulomb interaction is therefore screened by the geometric mean of the \emph{other two} principal permittivities. This has a counter-intuitive consequence: the interaction is \emph{least} screened along the axis of \emph{largest} permittivity. In rutile, two charges aligned along $c$ interact with $\epsilon_{\rm eff}=89$, whereas in the basal plane $\epsilon_{\rm eff}=\sqrt{89\times173}=124.1$ (Table~\ref{tab:eff}). Oppositely charged defect pairs are therefore more strongly bound when aligned along $c$.

For a hydrogenic (effective-mass) centre with an isotropic mass in a uniaxial host, the variational energy obtained with a spherical $1s$ trial function corresponds to an effective permittivity $\epsilon_D=\langle\epsilon_{\rm eff}^{-1}(\hat{\vR})\rangle^{-1}$, where the brackets denote an angular average:
\begin{equation}
\frac{1}{\epsilon_D}=\frac{1}{\sqrt{\epsilon_\perp\epsilon_\parallel}}\,\frac{\arcsin\sqrt{1-\epsilon_\perp/\epsilon_\parallel}}{\sqrt{1-\epsilon_\perp/\epsilon_\parallel}}\qquad(\epsilon_\parallel>\epsilon_\perp).\label{eq:donor}
\end{equation}
For rutile, $\epsilon_D=112.2$. This value is close to the geometric mean $(\epsilon_\perp^2\epsilon_\parallel)^{1/3}=111.1$ and noticeably smaller than the arithmetic mean $117.0$ that is often used. Equation~\eqref{eq:donor} only illustrates the effect of dielectric anisotropy: in rutile, mass anisotropy and polaronic effects must also be considered. The 2D analogue of Eq.~\eqref{eq:epseff} underlies the anisotropic Rytova--Keldysh interaction for excitons in phosphorene and related crystals \cite{Galiautdinov2019,Galiautdinov2026,Rodin2014,Prada2015}.

\begin{table}[!htbp]
\centering
\caption{Principal permittivities and direction-dependent Coulomb screening $\epsilon_{\rm eff}$ [Eq.~\eqref{eq:epseff}] along the principal axes for the tensors of Eq.~\eqref{eq:tensors}.}
\label{tab:eff}
\begin{tabular}{lcc}
\toprule
crystal & principal values $\epsilon_i$ & $\epsilon_{\rm eff}$ along the corresponding axes\\
\midrule
NaCl & $5.6,\;5.6,\;5.6$ & $5.6,\;5.6,\;5.6$\\
TiO$_2$ (rutile) & $89,\;89,\;173$ & $124.1,\;124.1,\;89.0$\\
CdWO$_4$ & $7.7,\;7.7,\;10.3$ & $8.9,\;8.9,\;7.7$\\
\bottomrule
\end{tabular}
\end{table}

%==========================================================================
\section{Green identities and boundary-value problems}\label{sec:bvp}

\subsection{Green identities and integral representation}
For two scalar fields $\varphi,\psi$ we apply the divergence theorem to $\mathbf{A}=\varphi\,\teps\nabla\psi$. Using $\nabla\cdot\mathbf{A}=\nabla\varphi\cdot\teps\nabla\psi+\varphi\nabla\cdot(\teps\nabla\psi)$ gives the anisotropic first identity. Antisymmetrising and using $\teps=\teps^{\top}$ gives the second identity:
\begin{align}
\int_\Omega\left[\nabla\varphi\cdot\teps\nabla\psi+\varphi\,\nabla\cdot(\teps\nabla\psi)\right]d^3r_0&=\oint_{S}\varphi\,\partial_\nu\psi\,dS_0,\label{eq:green1}\\
\int_\Omega\left[\varphi\,\nabla\cdot(\teps\nabla\psi)-\psi\,\nabla\cdot(\teps\nabla\varphi)\right]d^3r_0&=\oint_{S}\left(\varphi\,\partial_\nu\psi-\psi\,\partial_\nu\varphi\right)dS_0,\label{eq:green2}
\end{align}
where $\vn$ is the outward unit normal and $\partial_\nu\equiv\vn\cdot\teps\nabla_0$ is the \emph{conormal} derivative. It replaces the normal derivative of the isotropic theory, and $-\epsilon_0\partial_\nu\Phi=\vn\cdot\vD$ is the normal component of the displacement. The symmetry of $\teps$ is essential for the cancellation of the gradient terms in Eq.~\eqref{eq:green2}. Taking $\varphi=\Phi(\vr_0)$, $\psi=G(\vr,\vr_0)$ and using Eqs.~\eqref{eq:poisson}--\eqref{eq:Gdef} gives
\begin{equation}
\Phi(\vr)=\int_\Omega G(\vr,\vr_0)\rho(\vr_0)\,d^3r_0+\epsilon_0\oint_S\left[G(\vr,\vr_0)\,\partial_\nu\Phi(\vr_0)-\Phi(\vr_0)\,\partial_\nu G(\vr,\vr_0)\right]dS_0 .\label{eq:rep}
\end{equation}
The surface term involves the potential and the normal displacement $\vn\cdot\vD$ on $S$. As in the isotropic case, prescribing both overdetermines the problem. $G$ may be replaced by $G+F$, where $F$ satisfies $\nabla\cdot(\teps\nabla F)=0$ in $\Omega$ and represents charges outside $\Omega$, so that one of the two surface terms vanishes.

\subsection{Dirichlet and Neumann problems}
\emph{Dirichlet} (potential prescribed, e.g.\ electrodes): we choose $G_D$ with $G_D=0$ on $S$, so that
\begin{equation}
\Phi(\vr)=\int_\Omega G_D\rho\,d^3r_0-\epsilon_0\oint_S\Phi(\vr_0)\,\partial_\nu G_D(\vr,\vr_0)\,dS_0 .\label{eq:dirichlet}
\end{equation}
\emph{Neumann} (normal displacement prescribed, e.g.\ a boundary with a known surface charge, or an insulating boundary in the conduction analogue): applying Gauss' theorem to Eq.~\eqref{eq:Gdef} gives $\epsilon_0\oint_S\partial_\nu G\,dS_0=-1$. Hence $\partial_\nu G_N=0$ on $S$ is not allowed, and the simplest admissible choice is $\partial_\nu G_N=-1/(\epsilon_0 S)$, which gives
\begin{equation}
\Phi(\vr)=\langle\Phi\rangle_S+\int_\Omega G_N\rho\,d^3r_0+\epsilon_0\oint_S G_N(\vr,\vr_0)\,\partial_\nu\Phi(\vr_0)\,dS_0 ,\label{eq:neumann}
\end{equation}
where $\langle\Phi\rangle_S$ is the average potential on $S$. For exterior problems with $S\to\infty$ and $\Phi\to0$, the average vanishes and $\partial_\nu G_N=0$ can be used. Both $G_D$ and $G_N$ (the latter after adding a suitable function of $\vr_0$ \cite{Kim1993}) obey the reciprocity relation $G(\vr,\vr_0)=G(\vr_0,\vr)$. This follows from Eq.~\eqref{eq:green2} and therefore from the symmetry of $\teps$. In a gyrotropic medium, e.g.\ a conductor in a magnetic field with a non-symmetric $\bm\sigma$, reciprocity is lost, in agreement with the Onsager relations.

\subsection{Exact images for planar electrodes and insulating boundaries}\label{sec:mirror}
Consider the half-space $\vn\cdot\vr>0$ filled with the anisotropic medium, a charge $Q$ at $\vr_0$, and a planar boundary $\vn\cdot\vr=0$. We look for a point $\vr_0^*$ such that $\enorm{\vr-\vr_0}=\enorm{\vr-\vr_0^*}$ for every $\vr$ on the plane. A short calculation gives
\begin{equation}
\vr_0^{*}=\vr_0-2\,\frac{\vn\cdot\vr_0}{\vn\cdot\teps\vn}\;\teps\vn .\label{eq:emirror}
\end{equation}
This is the ``$\epsilon$-mirror'' of $\vr_0$: it lies at the same distance from the plane but on the other side, and it is reached by moving \emph{along $\teps\vn$} rather than along $\vn$. It follows that
\begin{equation}
G_D=G(\vr,\vr_0)-G(\vr,\vr_0^*),\qquad G_N^{\rm half}=G(\vr,\vr_0)+G(\vr,\vr_0^*)\label{eq:images}
\end{equation}
satisfy, \emph{for any orientation of the crystal axes}, $G_D=0$ (grounded electrode) and $\partial_\nu G_N^{\rm half}=0$ (boundary with no normal displacement) on the plane. This can be verified directly from Eq.~\eqref{eq:G}; we also checked it numerically for random positive-definite tensors (residuals at machine precision). The image lies directly below the charge only if $\vn$ is a principal axis. Otherwise it is displaced laterally by $-2(\vn\cdot\vr_0)\,\teps_{t n}/\epsilon_{nn}$, where $\teps_{tn}$ denotes the tangential components of $\teps\vn$. This generalises the results of Refs.~\cite{Lindell1997,Mele2001} for the two boundaries that are most common in practice. Examples are a charge in an anisotropic dielectric layer on a metallic back gate, or a scanning-probe tip on a crystal mounted on an electrode.

\emph{Conduction analogue and four-probe measurements.} Steady currents in a homogeneous conductor obey $\nabla\cdot(\bm\sigma\nabla\Phi)=0$ with $\mathbf{j}=-\bm\sigma\nabla\Phi$. All results above therefore apply with $\epsilon_0\teps\to\bm\sigma$; the same holds for heat conduction and diffusion. A current $I$ injected at a point of the free surface of a semi-infinite crystal is a source located on the boundary, so that $\vr_0^*=\vr_0$, and Eq.~\eqref{eq:images} doubles the bulk potential, $\Phi=I/[2\pi(\det\bm\sigma)^{1/2}|\vR|_{\sigma}]$, \emph{for any crystal face}. For a collinear four-probe array with equal spacing $s$ along the in-plane direction $\hat{\mathbf{u}}$, the measured resistance is therefore $V/I=\rho_{\rm app}(\hat{\mathbf{u}})/(2\pi s)$ with
\begin{equation}
\rho_{\rm app}(\hat{\mathbf{u}})=\frac{(\det\bm\rho)^{1/2}}{(\hat{\mathbf{u}}\cdot\bm\rho\,\hat{\mathbf{u}})^{1/2}},\qquad
\rho_{\rm app}(\hat{\mathbf{e}}_x)=\sqrt{\rho_y\rho_z}\ \ \text{(principal axes)},\label{eq:4probe}
\end{equation}
where $\bm\rho=\bm\sigma^{-1}$. A four-probe measurement along $x$ thus measures the geometric mean of the two \emph{transverse} resistivities. This is the semi-infinite counterpart of the Wasscher--Montgomery transformation used for anisotropic samples of finite size \cite{Montgomery1971}.

%==========================================================================
\section{Multipole expansion and metric-adapted moments}\label{sec:multipole}

\subsection{General expansion}
In the stretched coordinates of Eq.~\eqref{eq:stretch} the Green's function is Coulombic. The familiar spherical-harmonic expansion therefore applies with $\vs=\teps^{-1/2}\vr$, $s=\enorm{\vr}$ and $\hat{\vs}=\vs/s$. For a charge distribution confined to $\enorm{\vr_0}<\enorm{\vr}$,
\begin{equation}
\Phi(\vr)=\frac{1}{4\pi\epsilon_0(\dete)^{1/2}}\sum_{\ell=0}^{\infty}\sum_{m=-\ell}^{\ell}\frac{4\pi}{2\ell+1}\,q^{\epsilon}_{\ell m}\,\frac{Y_{\ell m}(\hat{\vs})}{s^{\ell+1}},\qquad
q^{\epsilon}_{\ell m}=\int\rho(\vr_0)\,s_0^{\ell}\,Y^{*}_{\ell m}(\hat{\vs}_0)\,d^3r_0 ,\label{eq:sph}
\end{equation}
with $\vs_0=\teps^{-1/2}\vr_0$. No Jacobian appears in $q^\epsilon_{\ell m}$ because $\rho\,d^3r_0$ is a charge element. The essential point is that the moments are computed with the \emph{stretched} positions $\vs_0$. At orders $\ell\ge2$ they therefore differ from the moments of the same charge distribution computed with the Euclidean metric.

\subsection{Cartesian form up to second order}
\begin{figure}[!htbp]
\centering
\includegraphics[width=0.66\linewidth]{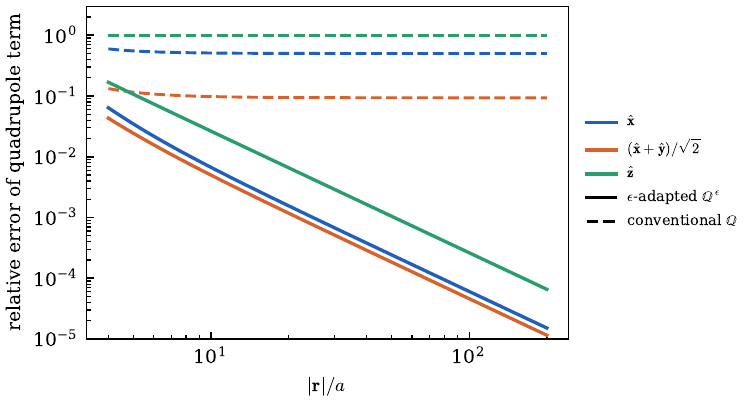}
\caption{Relative error of the second-order multipole approximation for a planar quadrupole of four charges $\pm Q$ at $(\pm a,\pm a,0)$ in CdWO$_4$, compared with the exact potential, along three directions. Solid lines: metric-adapted quadrupole $\Qe$ [Eq.~\eqref{eq:Qeps}]. Dashed lines: conventional traceless quadrupole $\mathbb{Q}$. Only $\Qe$ converges to the exact potential.}
\label{fig:conv}
\end{figure}

Expanding $\enorm{\vr-\vr_0}^{-1}$ to second order in $\vr_0$, using $(1+x)^{-1/2}=1-\tfrac12x+\tfrac38x^2+\mathcal{O}(x^3)$ \cite{jeffrey2007table}, gives
\begin{equation}
\frac{1}{\enorm{\vr-\vr_0}}=\frac{1}{\enorm{\vr}}+\frac{\vr\cdot\tepsi\vr_0}{\enorm{\vr}^3}
+\frac{3(\vr\cdot\tepsi\vr_0)^2-(\vr_0\cdot\tepsi\vr_0)\,\enorm{\vr}^2}{2\,\enorm{\vr}^5}+\cdots\label{eq:taylor}
\end{equation}
Inserting this in Eq.~\eqref{eq:Gdef} gives $\Phi=\Phi_M+\Phi_D+\Phi_Q+\cdots$ with
\begin{align}
\Phi_M(\vr)&=\frac{Q}{4\pi\epsilon_0(\dete)^{1/2}\enorm{\vr}},\qquad
\Phi_D(\vr)=\frac{\vr\cdot\tepsi\vp}{4\pi\epsilon_0(\dete)^{1/2}\enorm{\vr}^{3}},\label{eq:MD}\\
\Phi_Q(\vr)&=\frac{(\tepsi\vr)\cdot\Qe\,(\tepsi\vr)}{8\pi\epsilon_0(\dete)^{1/2}\enorm{\vr}^{5}},\label{eq:Q}
\end{align}
where $Q=\int\rho\,d^3r_0$ and $\vp=\int\rho\,\vr_0\,d^3r_0$ are the usual charge and dipole moment, and
\begin{equation}
\boxed{\;\Qe_{ij}=\int\rho(\vr_0)\left[3x_{0i}x_{0j}-\epsilon_{ij}\,(\vr_0\cdot\tepsi\vr_0)\right]d^3r_0\;},\qquad \tr\left(\tepsi\Qe\right)=0 .\label{eq:Qeps}
\end{equation}
In terms of the second-moment tensor $\mathsf{M}=\int\rho\,\vr_0\vr_0^{\top}d^3r_0$ and of the conventional traceless quadrupole tensor $\mathbb{Q}=3\mathsf{M}-\Id\,\tr\mathsf{M}$,
\begin{equation}
\Qe=3\mathsf{M}-\teps\,\tr(\tepsi\mathsf{M})=\mathbb{Q}+\Id\,\tr\mathsf{M}-\teps\,\tr(\tepsi\mathsf{M}).\label{eq:QeQ}
\end{equation}
The monopole and dipole moments are unchanged, but the quadrupole moment is not. It is ``traceless'' with respect to the metric $\tepsi$, and it depends on all six independent components of $\mathsf{M}$, whereas $\mathbb{Q}$ contains only five of them. Replacing $\Qe$ by $\mathbb{Q}$ in Eq.~\eqref{eq:Q}, i.e.\ assuming that the multipole moments keep their vacuum form, is therefore correct in general only for isotropic media, for which the last two terms of Eq.~\eqref{eq:QeQ} cancel. Two consequences are worth noting.
(i) A spherically symmetric charge distribution of radius $a$, such as a finite-size ion or a spherical charge cloud, has $\mathsf{M}=Qa^2\Id/3$ and therefore $\Qe=Qa^2\left[\Id-\teps\,\tr(\tepsi)/3\right]\ne0$. Its potential is \emph{not} that of a point charge beyond leading order. In rutile, $\Qe=Qa^2\,\mathrm{diag}(0.162,0.162,-0.629)$. Only distributions with ellipsoidal symmetry that matches the metric, $\rho=\rho(\vr_0\cdot\tepsi\vr_0)$, have $\Qe=0$.
(ii) For a planar quadrupole of four charges $\pm Q$ at $(\pm a,\pm a,0)$ in CdWO$_4$, $\mathbb{Q}$ is purely off-diagonal, whereas $\tr(\tepsi\mathsf{M})=-8Qa^2(\tepsi)_{xy}\neq0$. Figure~\ref{fig:conv} compares the exact potential of the four charges with the second-order term. With $\Qe$ the relative error decays as $(a/r)^2$, as expected, whereas with $\mathbb{Q}$ it saturates at 50\% along $x$ and 9\% along the diagonal, and reaches 100\% along $z$, where $\mathbb{Q}$ predicts a vanishing quadrupole field.

\begin{figure}[!htbp]
\centering
\includegraphics[width=0.92\linewidth]{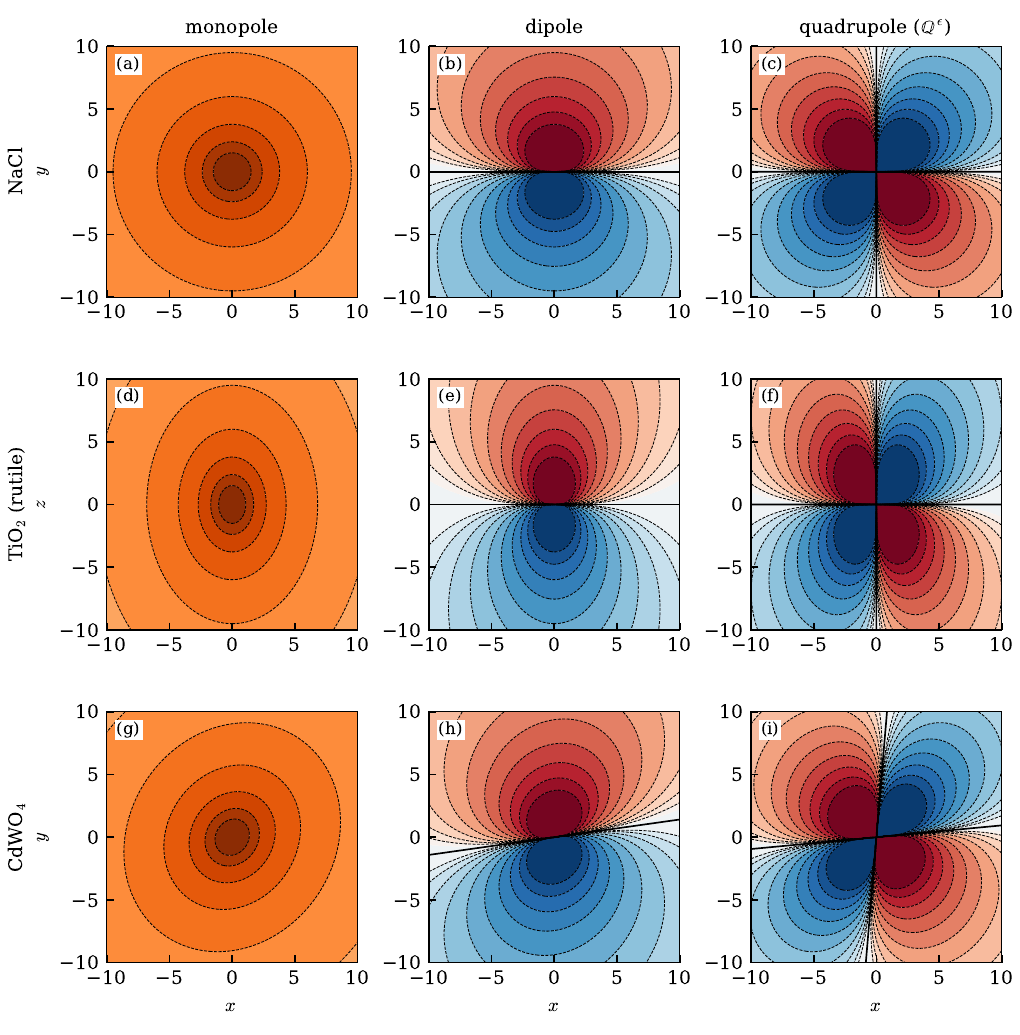}
\caption{Equipotentials of the monopole (left), dipole (centre) and quadrupole (right) terms, Eqs.~\eqref{eq:MD}--\eqref{eq:Qeps}, for NaCl ($xy$ plane), rutile TiO$_2$ ($xz$ plane, $z\parallel c$) and CdWO$_4$ ($xy$ plane), with the tensors of Eq.~\eqref{eq:tensors}. The dipole is oriented along the vertical axis of each panel. The quadrupole is the planar arrangement of charges $\pm Q$ at $(\pm a,\pm a)$ in the plotted plane. Colours use a signed logarithmic scale (red positive, blue negative). The solid black lines are the nodal lines. Lengths are in arbitrary units ($a=1$).}
\label{fig:multipoles}
\end{figure}

Figure~\ref{fig:multipoles} shows the monopole, dipole and quadrupole terms in the three crystals. In NaCl (a--c) the patterns are those of vacuum, rescaled by $\epsilon$. In rutile (d--f, $xz$ plane) the equipotentials are stretched along $c$ by $(\epsilon_\parallel/\epsilon_\perp)^{1/2}=1.39$. In CdWO$_4$ (g--i) the off-diagonal component $\epsilon_{xy}$ rotates the ellipses away from the crystal axes and tilts the nodal plane of the dipole. In the quadrupole term (i) it also moves the nodal lines; this effect is absent if the conventional $\mathbb{Q}$ is used. Outside a confined charge distribution, the equipotentials are thus distorted ellipsoids aligned with the principal axes of $\teps$, which are not, in general, the crystallographic axes of monoclinic and triclinic crystals \cite{nye1985physical,haussuhl2007physical}.

%==========================================================================
\section{Charges near the surface of an anisotropic dielectric}\label{sec:surface}
Scanning-probe techniques (Kelvin-probe and electrostatic-force microscopy), the mirror method of scanning electron microscopy, used to study charge trapping in insulators \cite{Damamme2008}, and the screening of 2D devices by their substrates \cite{Laturia2018,Deng2020} all involve a charge close to a dielectric surface. In contrast with the conductor and insulator boundaries of Sec.~\ref{sec:mirror}, the vacuum--dielectric interface of an anisotropic medium admits a point image only in special cases \cite{Mele2001,Lindell1997,Hanninen2004}. Here we give the general solution in a form that can be evaluated directly.

\subsection{Surface tensor}
Let the medium fill $z<0$ and vacuum fill $z>0$, and write the tensor in the surface frame as
\begin{equation}
\teps=\begin{pmatrix}\teps_t&\mathbf{e}\\ \mathbf{e}^\top&\epsilon_{zz}\end{pmatrix},\label{eq:block}
\end{equation}
with $\teps_t$ the $2\times2$ in-plane block and $\mathbf{e}=(\epsilon_{xz},\epsilon_{yz})$. A 2D Fourier component $e^{i\vk\cdot\bm\rho}$ of the potential, where $\bm\rho=(x,y)$ and $\vk=k(\cos\phi,\sin\phi)$, varies in the medium as $e^{\mu z}$ with $\epsilon_{zz}\mu^2+2i(\mathbf{e}\cdot\vk)\mu-\vk\cdot\teps_t\vk=0$. The surface response is entirely determined by the $2\times2$ \emph{surface tensor} and the associated surface permittivity
\begin{equation}
\mathsf{A}=\epsilon_{zz}\teps_t-\mathbf{e}\mathbf{e}^{\top},\qquad \det\mathsf{A}=\epsilon_{zz}\det\teps,\qquad
\eta(\phi)=\left(\hat{\vk}\cdot\mathsf{A}\hat{\vk}\right)^{1/2}.\label{eq:A}
\end{equation}
The ratio of normal displacement to potential at the surface is $\epsilon_0k\,\eta(\phi)$, compared with $\epsilon_0k$ in vacuum. The surface therefore behaves as an isotropic dielectric of permittivity $\eta$ only if $\mathsf{A}\propto\Id$. This is the case, for example, on the basal face of a uniaxial crystal, where $\eta=(\epsilon_\perp\epsilon_\parallel)^{1/2}$; in general $\eta_s=(\epsilon_{zz}\det\teps)^{1/4}$ when $\mathsf{A}\propto\Id$. Otherwise $\eta$ depends on the in-plane direction.

\subsection{External charge: image potential}
For a charge $Q$ at height $h$ above the surface, the reflected potential in vacuum follows from the 2D Fourier representation of $1/r$:
\begin{equation}
\Phi_{\rm refl}(\bm\rho,z)=\frac{Q}{4\pi\epsilon_0}\,\frac{1}{2\pi}\int_0^{2\pi}\!d\phi\;\frac{r(\phi)\,(z+h)}{(z+h)^2+(\bm\rho\cdot\hat{\vk})^2},\qquad r(\phi)=\frac{1-\eta(\phi)}{1+\eta(\phi)} .\label{eq:refl}
\end{equation}
If $\eta$ does not depend on $\phi$, this reduces to a point image $Q(1-\eta_s)/(1+\eta_s)$ at the mirror point. In general the image is a distribution (the ``anamorphosis'' of Ref.~\cite{Mele2001}). The self-energy that determines the image force on an ion, an adsorbate or an electron in front of the surface is $U(h)=Q^2\langle r\rangle/(16\pi\epsilon_0 h)$, where $\langle r\rangle$ is the average of $r(\phi)$ over $\phi$. For the CdWO$_4$ tensor of Eq.~\eqref{eq:tensors}, $\langle r\rangle=-0.789$ on the face normal to $x$ and $-0.785$ on the face normal to $z$. For high-permittivity rutile, $\langle r\rangle\simeq-0.98$ on all faces, and the image force is insensitive to anisotropy.

\subsection{Buried charge: apparent charge, depth and lateral position}
\begin{figure}[!htbp]
\centering
\includegraphics[width=\linewidth]{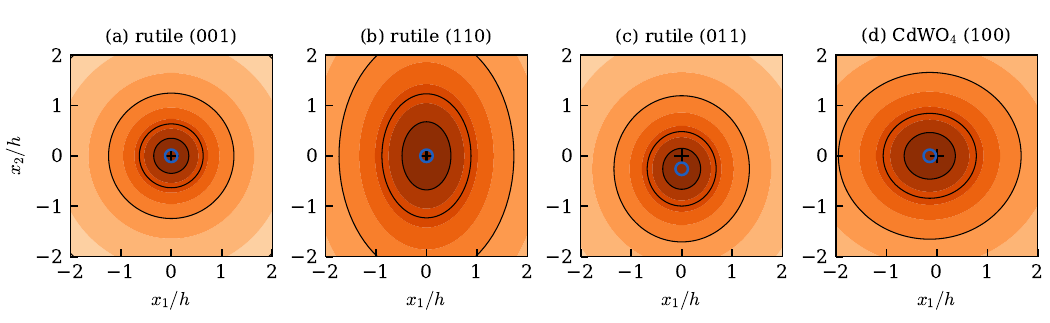}
\caption{Normalised potential on the surface ($z=0^+$) produced by a point charge buried at depth $h$, from Eq.~\eqref{eq:buried}. The cross marks the true lateral position of the charge and the open circle the apparent position $\bm\Delta$ of Eq.~\eqref{eq:happ}. (a) Rutile (001): circular pattern with apparent depth $0.72h$. (b) Rutile (110), with $c$ along the vertical axis: elliptical pattern elongated along $c$. (c) Rutile (011): the pattern is displaced by $0.26h$ along the projection of $c$. (d) CdWO$_4$, face normal to $x$: displacement of $0.14h$ caused by $\epsilon_{xy}$. Black lines are contours at 25, 50, 75 and 90\% of the maximum.}
\label{fig:buried}
\end{figure}

The situation is quite different for a charge $Q$ located at depth $h$ \emph{inside} the crystal. Matching the potential and the normal displacement at $z=0$ (Appendix~\ref{app:surface}) gives the potential in vacuum,
\begin{equation}
\Phi_{\rm vac}(\bm\rho,z)=\frac{Q}{4\pi\epsilon_0}\,\frac{1}{\pi}\int_0^{2\pi}\!d\phi\;\frac{1}{1+\eta(\phi)}\;
\frac{z+h_{\rm app}(\phi)}{\left[z+h_{\rm app}(\phi)\right]^2+\left[(\bm\rho-\bm\Delta)\cdot\hat{\vk}\right]^2},\label{eq:buried}
\end{equation}
\begin{equation}
h_{\rm app}(\phi)=h\,\frac{\eta(\phi)}{\epsilon_{zz}},\qquad \bm\Delta=h\,\frac{\mathbf{e}}{\epsilon_{zz}} .\label{eq:happ}
\end{equation}
Three effects can be read off directly.
(i) \emph{Apparent charge.} The charge seen from outside is reduced by $2/(1+\eta)$.
(ii) \emph{Apparent depth.} The charge appears at a depth $h_{\rm app}$ that differs from $h$ and that may depend on the in-plane direction, so that the surface potential is elliptical rather than circular. On the basal face of a uniaxial crystal, $h_{\rm app}=h(\epsilon_\perp/\epsilon_\parallel)^{1/2}$, and the vacuum potential is exactly that of a point charge $2Q/[1+(\epsilon_\perp\epsilon_\parallel)^{1/2}]$ at that depth.
(iii) \emph{Lateral shift.} When the surface normal is not a principal axis, the charge appears displaced laterally by $\bm\Delta$. Geometrically, the apparent position is the point where the line through the charge along the conjugate direction $\teps\vn$ meets the surface; this is the same direction that defines the $\epsilon$-mirror of Eq.~\eqref{eq:emirror}.
Table~\ref{tab:surface} and Fig.~\ref{fig:buried} give these quantities for several faces of rutile and for CdWO$_4$. For rutile, the same charge at the same true depth appears at $0.72h$ below the (001) face but at up to $1.39h$ below the (110) face (along the in-plane $c$ direction). On the (011) face, whose normal makes an angle of $32.8^\circ$ with $c$, it appears displaced by $0.26h$ along the projection of $c$. These are not small corrections. Depth profiles of trapped charge or dopants inferred from surface-potential maps with an isotropic model are therefore biased by factors of up to $\approx2$, and lateral positions by a fraction of the depth. For layered materials such as hBN, in which the out-of-plane permittivity is lower than the in-plane one \cite{Laturia2018}, Eq.~\eqref{eq:happ} predicts $h_{\rm app}>h$ on the basal face. Charged impurities in the substrate of a 2D device therefore appear both weaker and deeper than in an isotropic estimate.

\begin{table}[!htbp]
\centering
\caption{Surface quantities for a charge outside ($\langle r\rangle$) or buried at depth $h$ below several faces, from Eqs.~\eqref{eq:A}--\eqref{eq:happ} and the tensors of Eq.~\eqref{eq:tensors}. The lattice parameters $a=4.594$~\AA, $c=2.959$~\AA\ were used for rutile (011). Ranges reflect the dependence on the in-plane direction.}
\label{tab:surface}
\begin{tabular}{lcccccc}
\toprule
face & $\epsilon_{zz}$ & $\eta(\phi)$ & $h_{\rm app}/h$ & $|\bm\Delta|/h$ & $2/(1+\eta)$ & $\langle r\rangle$\\
\midrule
rutile (001) & 173.0 & 124.1 & 0.717 & 0 & 0.0160 & $-0.984$\\
rutile (110) & 89.0 & 89.0--124.1 & 1.000--1.394 & 0 & 0.0160--0.0222 & $-0.981$\\
rutile (011) & 148.4 & 114.9--124.1 & 0.775--0.836 & 0.258 & 0.0160--0.0173 & $-0.983$\\
CdWO$_4$, $\vn\parallel x$ & 8.50 & 8.09--8.91 & 0.952--1.048 & 0.141 & 0.202--0.220 & $-0.789$\\
CdWO$_4$, $\vn\parallel z$ & 7.70 & 7.70--8.91 & 1.000--1.157 & 0 & 0.202--0.230 & $-0.785$\\
\bottomrule
\end{tabular}
\end{table}

%==========================================================================
\section{Anisotropic inclusion in a uniform field (Laplace problem)}\label{sec:inclusion}
The response of a particle, pore or grain to a uniform applied field is the building block of dielectric mixing rules, of the interpretation of the dielectric response of ceramics and composites \cite{sihvola1999electromagnetic,Milton2002,Rodrigues2020}, and of dielectrophoresis. Consider an ellipsoidal inclusion with permittivity tensor $\teps_i$ in a host $\teps_h$, both arbitrary symmetric positive-definite tensors, with an applied uniform field $\vE_0$. In the coordinates $\vs=\teps_h^{-1/2}\vr$ the host becomes isotropic with unit permittivity. The inclusion becomes another ellipsoid with tensor $\tilde{\teps}_i=\teps_h^{-1/2}\teps_i\teps_h^{-1/2}$, and the uniform-field theorem for ellipsoids applies. Its depolarisation tensor $\mathsf{N}_s$ ($\tr\mathsf{N}_s=1$) is that of the \emph{transformed} ellipsoid \cite{Osborn1945}. Transforming back gives an exactly uniform internal field,
\begin{equation}
\vE_{\rm in}=\left[\Id+\mathsf{L}\,(\teps_i-\teps_h)\right]^{-1}\vE_0,\qquad \mathsf{L}=\teps_h^{-1/2}\,\mathsf{N}_s\,\teps_h^{-1/2},\qquad \tr(\teps_h\mathsf{L})=1 .\label{eq:Ein}
\end{equation}
The perturbation of the field outside the inclusion is, at large distance, the field of the dipole $\vp=\epsilon_0 V(\teps_i-\teps_h)\vE_{\rm in}$ in the host, given by $\Phi_D$ of Eq.~\eqref{eq:MD} with $\teps=\teps_h$. This defines the polarisability $\bm\alpha=\epsilon_0V(\teps_i-\teps_h)\left[\Id+\mathsf{L}(\teps_i-\teps_h)\right]^{-1}$. Two cases of practical interest follow immediately.

\emph{Anisotropic sphere in an isotropic host} ($\teps_h=\epsilon_h\Id$, $\mathsf{L}=\Id/(3\epsilon_h)$):
\begin{equation}
\vE_{\rm in}=3\epsilon_h\left(\teps_i+2\epsilon_h\Id\right)^{-1}\vE_0,\qquad \bm\alpha=4\pi\epsilon_0\epsilon_hR^3\,(\teps_i-\epsilon_h\Id)(\teps_i+2\epsilon_h\Id)^{-1}.\label{eq:sphere1}
\end{equation}
The internal field is not parallel to $\vE_0$ unless $\vE_0$ is along a principal axis. For a rutile sphere in vacuum, the internal field is $0.033E_0$ for $\vE_0\perp c$ and $0.017E_0$ for $\vE_0\parallel c$.

\begin{figure}[!htbp]
\centering
\includegraphics[width=\linewidth]{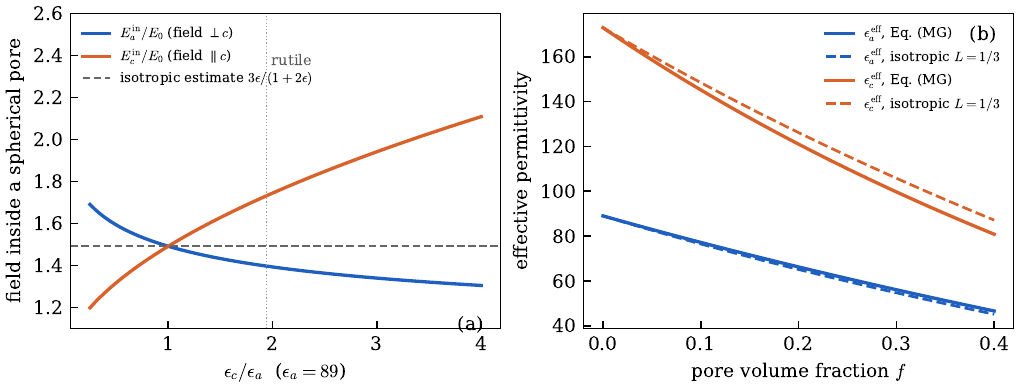}
\caption{Spherical pore ($\epsilon_i=1$) in a uniaxial host. (a) Field inside the pore for $\vE_0\perp c$ and $\vE_0\parallel c$ as a function of the anisotropy ratio, with $\epsilon_a=89$, from Eq.~\eqref{eq:Ein}. The dashed line is the isotropic estimate. The dotted line marks rutile. (b) Maxwell Garnett effective permittivity of porous textured rutile, Eq.~\eqref{eq:MG} (solid lines), compared with the component-wise isotropic formula with $L=1/3$ (dashed lines).}
\label{fig:inclusion}
\end{figure}

\emph{Isotropic sphere or spherical pore in an anisotropic host.} A sphere of radius $R$ maps onto an ellipsoid with semi-axes $R/\epsilon_k^{1/2}$ along the principal axes of $\teps_h$. Hence $\mathsf{L}$ is diagonal in that frame with $L_k=N_k/\epsilon_k$, where the $N_k$ are the depolarisation factors of that ellipsoid. It is therefore \emph{not} equal to $\Id/(3\epsilon_k)$, as a component-by-component use of the isotropic formula would suggest. For a spherical pore in rutile, the transformed ellipsoid is oblate with $N_a=0.287$ and $N_c=0.425$. The field inside the pore is then $1.40E_0$ ($\vE_0\perp c$) and $1.73E_0$ ($\vE_0\parallel c$), whereas the component-wise isotropic estimate $3\epsilon_k/(1+2\epsilon_k)$ gives $1.49E_0$ in both cases. The difference reaches 16\% (Fig.~\ref{fig:inclusion}a) and is relevant, for example, for the onset of partial discharges in porous high-permittivity ceramics.

\emph{Effective permittivity.} For a dilute or moderately concentrated suspension of identical, aligned inclusions at volume fraction $f$, averaging $\vD$ and $\vE$ with $\vE_{\rm in}$ from Eq.~\eqref{eq:Ein} gives the anisotropic Maxwell Garnett formula \cite{Levy1997,Giordano2006,sihvola1999electromagnetic}
\begin{equation}
\teps_{\rm eff}=\teps_h+f\,(\teps_i-\teps_h)\left[\Id+(1-f)\,\mathsf{L}\,(\teps_i-\teps_h)\right]^{-1}.\label{eq:MG}
\end{equation}
For textured rutile with spherical pores, Eq.~\eqref{eq:MG} gives $\epsilon^{\rm eff}_a=66.2$ and $\epsilon^{\rm eff}_c=121.0$ at $f=0.2$, compared with $65.1$ and $126.2$ from the component-wise isotropic formula. The error of the latter grows with porosity and anisotropy, reaching about 6\% for $\epsilon^{\rm eff}_c$ at $f=0.3$ (Fig.~\ref{fig:inclusion}b). Equation~\eqref{eq:MG} is the quantity that should be compared with the components of the dielectric tensor of oriented compacts measured, for example, by infrared reflectivity or ellipsometry \cite{Rodrigues2020,Schubert1998}. For randomly oriented anisotropic grains, Eq.~\eqref{eq:Ein} remains the building block, but $\teps_{\rm eff}$ requires an orientational average or a self-consistent (Bruggeman-type) scheme \cite{Levy1997,Milton2002}.

%==========================================================================
\section{Conclusions}\label{sec:conclusion}
Electrostatics in a homogeneous, linear, non-magnetic anisotropic dielectric is governed by the metric $\vr\cdot\tepsi\vr$ and by the factor $(\dete)^{-1/2}$. With the conormal derivative $\vn\cdot\teps\nabla$ replacing the normal derivative, the Green identities and the Dirichlet and Neumann representations keep their familiar form, and reciprocity follows from the symmetry of $\teps$. The multipole expansion has the same structure as in vacuum only up to the dipole. From the quadrupole order on, the moments must be computed in the stretched coordinates, and the metric-adapted tensor $\Qe$ of Eq.~\eqref{eq:Qeps} replaces the conventional traceless quadrupole, which does not give the correct far field. Several consequences can be tested experimentally or used in data analysis. The Coulomb interaction is least screened along the axis of largest permittivity. Planar electrodes and insulating boundaries admit an exact image at the $\epsilon$-mirror point for any crystal orientation, which gives a closed-form four-probe resistance for anisotropic conductors. Buried charges are seen from outside with an apparent charge, depth and lateral position that depend on the crystal face, with corrections of up to a factor of two in rutile. Finally, the Laplace problem of an anisotropic ellipsoid in an anisotropic host has a closed-form solution, which quantifies the error of component-wise isotropic estimates of local fields and effective permittivities in porous textured ceramics. Natural extensions are layered geometries with several anisotropic films, for which Eq.~\eqref{eq:A} provides the surface tensor of each interface, and the frequency-dependent regime near polar-phonon resonances, where $\teps(\omega)$ becomes hyperbolic.

%==========================================================================
\appendix
\section{Derivation of Eqs.~\eqref{eq:buried}--\eqref{eq:happ}}\label{app:surface}
For a 2D Fourier component $e^{i\vk\cdot\bm\rho}$, the roots of $\epsilon_{zz}\mu^2+2i(\mathbf{e}\cdot\vk)\mu-\vk\cdot\teps_t\vk=0$ are $\mu_\pm=\left[-i\,\mathbf{e}\cdot\vk\pm k\eta(\phi)\right]/\epsilon_{zz}$. The quantity under the square root reduces to $\vk\cdot\mathsf{A}\vk$ with $\mathsf{A}$ defined in Eq.~\eqref{eq:A}. For a unit charge at $z=-h$, the solution that is regular at $z\to-\infty$ is $\propto e^{\mu_+(z+h)}$ below the charge and contains $e^{\mu_\pm(z+h)}$ between the charge and the surface. In vacuum it is $\propto e^{-kz}$. The normal displacement associated with $e^{\mu z}$ is $-\epsilon_0(\epsilon_{zz}\mu+i\,\mathbf{e}\cdot\vk)$, which equals $\mp\epsilon_0k\eta$ for $\mu_\pm$. The jump condition at the charge fixes the amplitude of the direct wave to $1/(2\epsilon_0k\eta)$. Continuity of $\Phi$ and $\vn\cdot\vD$ at $z=0$ then gives the transmission factor $2\eta/(1+\eta)$, so that the vacuum amplitude is
\[
\tilde\Phi_{\rm vac}(\vk,z)=\frac{1}{\epsilon_0 k\,(1+\eta)}\,e^{-i\vk\cdot\mathbf{e}h/\epsilon_{zz}}\,e^{-k\eta h/\epsilon_{zz}}\,e^{-kz}.
\]
Integrating over $k$ in polar coordinates gives Eq.~\eqref{eq:buried}. The phase factor produces the shift $\bm\Delta$ and the decay rate produces $h_{\rm app}$. We verified this amplitude against a direct numerical solution of the four matching conditions for the rutile and CdWO$_4$ faces of Table~\ref{tab:surface}. We also checked that Eq.~\eqref{eq:buried} reduces to $2Q/[4\pi\epsilon_0(1+\epsilon)r]$ in the isotropic limit, and that the maximum of the surface potential on rutile (011) lies at $\bm\Delta$. Equation~\eqref{eq:refl} is obtained in the same way, with reflection factor $(1-\eta)/(1+\eta)$ and the 2D representation $1/r=(2\pi)^{-1}\int d^2k\,k^{-1}e^{i\vk\cdot\bm\rho-k|z|}$.

%==========================================================================
\section*{Author declarations}
\noindent\textbf{Conflict of interest.} The author has no conflicts to disclose.\\
\textbf{Author contributions.} J.E.F.S.R.: conceptualization, methodology, formal analysis, software, validation, visualization, writing (original draft, review and editing).\\
\textbf{Data availability.} The scripts used to generate the figures and the numerical checks reported in this article are available from the author upon reasonable request.

\bibliographystyle{unsrtnat}
\bibliography{references}

\end{document}